\documentclass[twocolumn]{aastex701}
\usepackage{amsthm,amsmath,amsfonts,amssymb}

\usepackage{mathtools,breqn,comment,graphicx}
\usepackage{xcolor,epstopdf}
\usepackage[caption=false]{subfig}

\theoremstyle{plain}

\newtheorem{conjecture}{Conjecture}

\theoremstyle{definition}

\theoremstyle{remark}
\newtheorem{remark}{Remark}

\usepackage{graphicx,psfrag,epsf}
\usepackage{enumerate}

\usepackage{color,xcolor,nicefrac}

\def\cstat{${C}$ statistic}
\def\c{$C$}

\def\1es{1ES~1553+113}
\def\dof{degrees--of--freedom}

\def\cmin{$C_{\mathrm{min}}$}
\def\cmineq{C_{\mathrm{min}}}
\def\ctrue{$C_{\mathrm{true}}$}
\def\cstar{$C^{\star}$}
\def\cstareq{C^{\star}}
\def\cstat{${C}$ statistic}
\def\wstat{${W}$ statistic}

\def\wmin{$W_{\mathrm{min}}$}
\def\wmineq{W_{\mathrm{min}}}
\def\df{\mathrm{df}}

\newcommand\Var{\text{Var}}
\newcommand\Cov{\text{Cov}}

\newcommand\E{\text{E}}

\def\chimin{$\chi^2_{\mathrm{min}}$}
\def\Poiss{\mathrm{Poisson}}

\def\ml{maximum--likelihood}

\def\L{$\mathcal{L}$}

\def\E{\mathrm{E}}

\def\gof{goodness--of--fit}

\definecolor{darkgreen}{rgb}{0.0, 0.5, 0.0}

\colorlet{blue}{black}
\newcommand{\XL}[1]{\textbf{\color{blue}XL: #1}}

\shorttitle{Poisson regression with background}
\shortauthors{Bonamente}
\begin{document}

\title{A comparison of methods for Poisson regression in the presence of background}

\author[0000-0002-8597-9742]{Massimiliano Bonamente}\affiliation{Department of Physics and Astronomy, University of Alabama in Huntsville, Huntsville, AL 35899}\affiliation{Department of Mathematical Sciences, University of Alabama in Huntsville, Huntsville, AL 35899}
\email[show]{bonamem@uah.edu}  

\author[0000-0002-3869-7996]{Vinay Kashyap}\affiliation{Center for Astrophysics $|$ Harvard \& Smithsonian}
\email{vkashyap@cfa.harvard.edu}

\author{Xiaoli Li}\affiliation{Department of Statistics, University of Chicago, Chicago, IL 60637}
\email{xiaolil@uchicago.edu}

\author{Jelle de Plaa}\affiliation{SRON Space Research Organisation Netherlands, Leiden, The Netherlands}\email{j.de.plaa@sron.nl}

\begin{abstract}
This paper provides a statistical analysis of three common methods of regression for Poisson data
in the presence of Poisson background, namely the joint fit with two parametric models for the source and the background, the use of a non-parametric model for the background known as the \emph{wstat} method, and the regression with a fixed background. The non-parametric background method, which is a popular method for
spectral data, is found to be significantly biased, especially in the low-count and background-dominated regimes. Similar conclusions apply to the fixed-background regression. The joint-fit method, on the other hand, simultaneously affords reliable hypothesis testing by means of the usual \emph{Cash} statistic and unbiased reconstruction of source parameters. We also investigate the effect of non-parametric regression on the number of effective degrees of freedom by means of the Efron degree of freedom function. We find that the \emph{wstat} 
method adds a significantly larger number of degrees of freedom, compared to the number of free parameters in the source model. The other
two methods have a number of degrees of freedom consistent with the number of adjustable parameters, at least for the simple models investigated in this paper. 
\end{abstract}

\keywords{Astrostatistics(1882);  Maximum likelihood estimation(1901); Poisson distribution(1898); Parametric hypothesis tests(1904); Nonparametric hypothesis tests(1920)}

\section{Introduction}
\setcounter{footnote}{0}
Maximum-likelihood Poisson regression is a key
tool for the analysis of event data in astronomy and across
the sciences \citep[e.g.,][]{cameron2013}. Methods
for the assessment of the goodness-of-fit were
established and popularized for astronomy by \cite{cash1976, cash1979}, leading
to his namesake \gof\ statistic 
that is usually known as the Poisson \emph{deviance} in other fields \citep[e.g.,][]{bishop1975}.

It is a common occurrence that spectra or light curves from astronomical sources are obtained
from a region that comprises both the source of interest and a background \citep[e.g.,][]{li1983},
using a secondary source-free region to estimate the background. The process
of background subtraction introduces a significant complication to the analysis
of the data, given that the difference between two Poisson variables does not retain the 
same distribution. In the large-count limit, this limitation is usually overcome
by approximating the data with a Gaussian distribution, and using the Pearson or Neyman chi-squared
statistics for the regression and subsequent \gof\ assessment \citep[see][for a review]{greenwood1996}.

For low-count data, which are especially common in high-energy astrophysics but also in other
fields, it is necessary to  devise methods of
regression and \gof\ that retain 
consistency with
the Poisson distribution of both data and background. In particular,
a popular method of regression in the presence of Poisson background for spectral data
is the one described in \cite{vianello2018} and \cite{xspecManual}, and known in the high-energy astrophysics community as the \emph{wstat} or $W$ statistic method.
This method consists in the use of a non-parametric background model for the estimation
of the source parameters, and is
in fact available in all major spectral fitting packages \citep[e.g., Sherpa, SPEX and XSPEC,][]{kaastra1996,arnaud1996}.\footnote{There are various and sometimes conflicting uses of the terms \emph{Cash}, \emph{cstat}
and \emph{wstat} in the astronomical literature and software. A discussion is provided
at the end of Sec.~\ref{sec:Data} after the relevant statistics have been defined.}
Given its simplicity and purported ability to estimate best-fit parameters and the \gof\ of the regression, this is
a convenient method of analysis for spectral data \citep[e.g., as in][]{spence2024, bonamente2025c}. 

The main motivation of this paper is to presents a statistical analysis of this popular method 
of regression, and to provide a comparison with two possible alternatives,
viz. the use of a constant background and the joint fit of source and background data{\color{blue}, which are
the other two most popular methods of background fitting in astrophysics}.
The goal is to investigate the distributions of the \gof\ statistics that derive from these methods, and
to determine the biases introduced in the estimation of source parameters, by means of numerical simulations
and analytical considerations. {\color{blue} Similar methods of background modeling apply to related disciplines such as particle physics, where both parametric
and non-parametric methods of background subtractions are commonly used \citep[e.g., see Appendix 1 of][]{lyons2008, Dauncey_2015, chisholm2022}.}

This paper is structured as follows: Sec.~\ref{sec:Data} presents the data model {\color{blue} and the three methods of background modeling under investigation}, in 
Sec.~\ref{sec:statistics} we provide the theoretical background of the fit statistics,
Sec.~\ref{sec:MonteCarlo} presents the results of Monte Carlo simulations and in Sec.~\ref{sec:discussion}  we discuss the results and present our conclusions.

\section{Data and parameter estimation}
\label{sec:Data}

This section presents the statistical model for the data and three of the most
commons methods for estimating the parameters of interest {\color{blue} in astrophysical applications}.

\subsection{The data model}
\label{sec:PoissonData}

It is assumed that there are two independent observations: one of a source region (S), and 
another one of a background region (B). Each observation represents a different spatial
region and different time intervals, 
with $t_S$ and $t_B$ representing a suitable product of
area and time for the two regions.
Each region
yields
data  of the type $(x_i, y_i)$, for $ i=1,\dots,N$ independent
 Poisson measurements $y_i$ at different values of independent variables $x_i$, same as
 in \cite{bonamente2025a, bonamente2025b}.
The  $x$ variable, with its $x_i$ fixed positions,
represents the predictor or independent variable (i.e., wavelength or energy for spectra, or time respectively for light curves), and they are the same for the two regions.
The response or dependent variable $y_i$ is respectively indicated as $S_i$ and $B_i$ for the two
regions.

The two datasets are assumed to be drawn from two independent parent Poisson distributions, respectively
\begin{equation}
    \begin{cases}
        S_i \sim \Poiss(\lambda_i)\quad \text{with } \lambda_i=(\mu_i(\theta)+b_i)\cdot t_S \\
        B_i \sim \Poiss(\beta_i) \quad \text{with } \beta_i=b_i \cdot t_B
        \end{cases}
\label{eq:dataModel}
\end{equation}
where $t_S$ and $t_B$ represents the known size and exposure time of the two regions,
$\mu_i=\mu(x_i;\theta)$ the model of interest for the background-subtracted signal,
and $b_i$ the model for the background. 

The main purpose of the regression is to estimate 
$\theta=(\theta_1,\dots,\theta_m)$, which is a set of $m$ adjustable parameters for
the model of interest, e.g., a power-law or a thermal emission model for spectral data, etc..
The data generating process for the background may also be described by another set of $a$
ancillary parameters for the parent background model $b_i$, say $\phi=(\phi_1,\dots,\phi_{a})$.

Notice that $\mu_i$ and $b_i$ can be considered parent models per unit time and area for, respectively, the
background-subtracted source signal and the background. We chose to retain the Latin-alphabet 
notation for the background model ($b_i$), for consistency with \cite{vianello2018}.  Since these data-generating processes 
are assumed to have a constant rate throughout the observation period and areas,
same conclusions would apply by using a model for the total number of counts. 

\subsection{The joint fit method}
\label{sec:jointFit}
The joint likelihood with the data model \eqref{eq:dataModel} is simply the product of
the Poisson likelihoods of the source and background regions, 
\begin{equation}
    \begin{aligned}
        \mathcal{L}(\theta,\phi)& = \mathcal{L}_S(\theta, \phi)  \cdot \mathcal{L}_B(\phi)\\
        & = \prod_{i=1}^N \dfrac{e^{-\lambda_i}{\lambda_i^{S_i}}}{S_i!} \cdot  \prod_{i=1}^N \dfrac{e^{-\beta_i}{\beta_i^{B_i}}}{B_i!}.
    \end{aligned}
    \label{eq:JointPoissonLikelihood}
\end{equation}
 The associated \gof\ statistic is the joint \cmin\
statistic, defined as usual via
\begin{equation}
    \cmineq=-2 (\ln \mathcal{L}(\hat{\theta},\hat{\phi}) - \ln \mathcal{L}(y,y))
    \label{eq:cminLR}
\end{equation}
where $\ln \mathcal{L}(y,y)$ represents the maximum achievable likelihood \citep[see, e.g.][]{bonamente2025a} and $\hat{\theta}$ and $\hat{\phi}$ are the \ml\ estimate of the parameters. This
statistic is usually referred to as the Poisson \emph{deviance} in statistical literature \citep[e.g.,][]{bishop1975, cameron1986} and as the \emph{Cash} statistics in the astronomical literature \citep{cash1976,cash1979}.
In this application, the statistic takes the form 
\begin{equation}
\begin{aligned}
    \cmineq &=  -2 \sum_{i=1}^N (S_i - (\hat{\mu}_i+\hat{b}_i) t_S) + S_i \ln \left( \dfrac{(\hat{\mu}_i+\hat{b}_i) t_S}{S_i} \right) \\
    & -2 \sum_{i=1}^N (B_i - \hat{b}_i t_B) 
     +B_i \ln \left( \dfrac{\hat{b}_i t_B}{B_i}\right) \\ 
     &\equiv  \sum_{i=1}^N C_{\mathrm{min},i}(S_i) + \sum_{i=1}^N C_{\mathrm{min},i}(B_i)
    \end{aligned}
    \label{eq:cmin}
\end{equation}
and it is the sum of the \cmin\ statistics  for the two independent data sets $S_i$ and $B_i$,
fitted simultaneously to obtain the two parameter sets $\hat{\theta}$ and $\hat{\phi}$.

The method of analysis that uses this joint likelihood is the most appropriate, given
the assumed data model and the chosen \ml\ method of regression. However, it requires that the background be also modeled with an additional
set of parameters, say $b_i=b(x_i;\phi)$, so that $\hat{b}_i=b(x_i;\hat{\phi})$. This is a burden for those
applications where a simple model cannot be found, as in certain astrophysical contexts \citep[e.g.,][]{nevalainen2005}, because it leads to a more complex fitting process.

\subsection{The non-parametric background method}
\label{sec:wstatData}
\cite{vianello2018} and Appendix~B in \cite{xspecManual} 
discuss a method of regression that can be used when a parametric model for the background is not available. The method consists in the use of a non-parametric or step-wise constant model for
the background, so that, within the usual parametric framework, one can set
$b=(b_1,\dots,b_N)=(\phi_1,\dots,\phi_N)=\phi$ as the model
for the background.  Within the parametric setting of Sec.~\ref{sec:PoissonData}, this is therefore an $N$-parameter model, although it is in reality a non-parametric model. For a fixed set of source-parameter values $\theta$, the value of $\hat{b}_i$ in each bin is obtained as a function of the parameters 
of interest $\theta$ by 
\begin{equation}
    \dfrac{\partial \mathcal{L}(\theta,b)}{\partial b_i}=0\quad \text{for } i=1,\dots, N.
    \label{eq:biML}
\end{equation}
These conditions lead to $N$ functions $\hat{b}_i(\theta)$ that are given by
\begin{equation}
    \begin{cases}
        \hat{b}_i(\theta)= \dfrac{1}{2} \left( \dfrac{S_i+B_i}{t_S+t_B} - \mu_i(\theta) + \sqrt{\Delta}\right), \text{with}\\[10pt]
         \Delta=\left( \mu_i(\theta) - \dfrac{S_i+B_i}{t_S+t_B}\right)^2 + \dfrac{4 \cdot B_i \cdot \mu_i(\theta)}{t_S+t_B}.
    \end{cases}
\label{eq:biHat}
\end{equation}

This method is therefore non-linear, in that the
fitted values are non-linear functions of the data. Such non linearity prevents the
use of methods of \gof\ assessment that are available to linear non-parametric models \citep[e.g., see discussion in][and references therein]{hidalgo2018, azzalini1993}. This issue will be investigated in detail in Sec.~\ref{sec:statisticsDistributions}.

When the background region has no counts in a given bin, $B_i=0$, Equations~\eqref{eq:biHat} lead to the following
estimate:
\begin{equation}
\begin{aligned}
    &\hat{b}_i(\theta) = \dfrac{1}{2} \left(\dfrac{S_i}{t_S+t_B} - \mu_i(\theta) + \left| \mu_i(\theta)-\dfrac{S_i}{t_S+t_B} \right| \right)\\
    & =\begin{cases}
        0 \quad \text{if} \quad  \mu_i(\theta)>\dfrac{S_i}{t_S+t_B}\\
        \dfrac{S_i}{t_S+t_B} - \mu_i(\theta)\geq 0 \quad \text{if} \quad  \mu_i(\theta) \leq \dfrac{S_i}{t_S+t_B}
    \end{cases}
    \end{aligned}
    \label{eq:biZero}
\end{equation}
as also discussed in \cite{xspecManual}
This means that for certain low-count datasets there may be many bins where the estimated background model
is $\hat{b}_i=0$. Such estimate is outside of the range that is allowable for a Poisson mean, which needs to
be strictly positive. Implications of this peculiarity of the \wstat\ method are discussed in Sec.~\ref{sec:MonteCarlo}.

The estimated parameters $\hat{b}_i(\theta)$ according to \eqref{eq:biHat} are known as \emph{restricted} \ml\ estimates or rMLEs, i.e., they maximize the likelihood for a fixed value of
the interesting parameters $\theta$
\citep[for a textbook review, see e.g.][]{pawitan2001}. Their functional forms can then 
be used in the joint likelihood to obtain the profile likelihood 
\begin{equation}
    \mathcal{L}_{\theta}(\theta)=\underset{b}{\text{max}} \;\mathcal{L}(\theta,b) =  \mathcal{L}(\theta,\hat{b}(\theta))
    \label{eq:profileLikelihood}
\end{equation}
where $(\theta,b)$ is the full parameter vector for the joint likelihood, with $b=(b_1,\dots,b_N)$
the non-parametric model for the background data, and the maximization is done at constant $\theta$. 
This profile likelihood is a likelihood in the ordinary sense for the unknown interesting parameters $\theta$.

Finally, the \ml\ estimate for the parameters of interest is obtained by usual maximization of this
profile likelihood. In turn, this leads to a \gof\ statistic that
is formally the same as \eqref{eq:cmin} and where $\hat{b}_i=\hat{b}_i(\hat{\theta})$, i.e.,
 using the function in \eqref{eq:biHat} evaluated at the global MLE $\hat{\theta}$:
\begin{equation}
    \begin{aligned}
    &\wmineq= -2 \sum_{i=1}^N (S_i - \left(\hat{\mu}_i+\hat{b}_i(\hat{\theta})\right) t_S) + (B_i - \hat{b}_i(\hat{\theta}) t_B)\\
    &+  S_i \ln \left( \dfrac{\left(\hat{\mu}_i+\hat{b}_i(\hat{\theta})\right) t_S}{S_i} \right) +B_i \ln \left( \dfrac{\hat{b}_i(\hat{\theta}) t_B}{B_i}\right).
    \end{aligned}
    \label{eq:wmin}
\end{equation}
This is the \emph{wstat} statistic discussed by \cite{vianello2018} and \cite{xspecManual} and commonly used in high-energy astrophysics data analysis software \citep[e.g.,][]{kaastra1996, arnaud1996}.
Contrary to the simultaneous estimation of all parameters with a joint fit to the data, first both datasets $S_i$ and $B_i$ are used to obtain the rMLE $\hat{b}_i(\theta)$, and then again both datasets 
are used to estimate the interesting parameters $\hat{\theta}$. This method of regression
will cause \wmin\ to be systematically
smaller than the fixed-background \cmin, as some of the variance (from the data $S_i$) is already explained by the fitted background parameters $\hat{b}_i$ (see also remark~\ref{remark:Wmin<Cmin} below).

\subsection{The fixed background method}

A more crude method of regression consists of simply ignoring the statistical model for the 
background data $B_i$, and assume that the background data are fixed. In practice, this consists
of assuming that the source data now have a model
\begin{equation}
    S_i \sim \Poiss(\mu_i(\theta)+B_i/t_B)\cdot t_S
    \label{eq:SiFixed}
\end{equation}
which corresponds to rescaling the measured background $B_i$ by the ratio
$t_S/t_B$ when applied to the source region.

This method leads to the usual \cmin\ \gof\ statistic for the source data alone, i.e.,
\begin{equation}
\begin{aligned}
    \cmineq=& -2 \sum_{i=1}^N (S_i - (\hat{\mu}_i+B_i/t_B) \cdot t_S) \\
    + & S_i \ln \left( \dfrac{(\hat{\mu}_i+B_i/t_B) \cdot t_S}{S_i} \right)
    \end{aligned}
    \label{eq:cminCB}
\end{equation}
Of course, the best-fit model parameters for this method differ from those for the
joint fit and the non-parametric background model, although they are indicated with 
the same symbols in \eqref{eq:cmin} and in \eqref{eq:cminCB}. Note how a key assumption is that
the measured background is the true background.

{
\subsection{A note on terminology for Poisson log-likelihoods}
\label{sec:terminology}
\cite{cash1979} introduced the Poisson log-likelihood
for parameter estimation in astronomy, by defining the $C$ statistic as
\begin{equation}
    C = -2 \ln  \mathcal{L} = 2 \sum_{i=1}^N \mu_i - y_i \ln \mu_i + \ln y_i!,
    \label{eq:CashCstat}
\end{equation}
where as always $y_i$ are detected counts and $\mu_i$ are model predictions, see their Eq.~(3).~\footnote{It also introduces a $\Delta C$ statistic that is useful for parameter estimation, by use of
Wilks' theorem, see their Eq.~(4); this is the same $\Delta C$ statistic discussed in \cite{bonamente2025b}, which
is not discussed in the present paper. }
It also correctly
identifies its asymptotic chi-squared distribution by subtraction of another data-dependent term,
\begin{equation}
\begin{aligned}
   & \hat{C} - 2\sum_{i=1}^N (y_i \ln y_i + \ln y_i!) = 2 \sum_{i=1}^N (\hat{\mu}_i - y_i) + y_i \ln \dfrac{y_i}{\hat{\mu}_i} =\\
   &\simeq
   \sum_{i=1}^N \left(\dfrac{y_i - \hat{\mu}_i}{\hat{\mu}_i}\right)^2 \left[1+O(1/\sqrt{y_i})\right],
   \label{eq:CashDistr}
   \end{aligned}
\end{equation}
without making use of the Wilks' likelihood-ratio theorem \citep{wilks1938}.

\cite{baker1984} provide the first expression of the `Poisson likelihood chi-square' as the \cmin\ statistic in the form 
currently used, i.e., as the left hand side of Eq.~\eqref{eq:CashDistr},
\begin{equation}
    \cmineq = 2 \sum_{i=1}^N (\hat{\mu}_i - y_i) + y_i \ln \dfrac{y_i}{\hat{\mu}_i} = -2 \ln \dfrac{\mathcal{L}(\hat{\theta})}{\mathcal{L}(y)}.
    \label{eq:CminBaker}
\end{equation}
They derive this statistic as a likelihood ratio between the maximum likelihood $\mathcal{L}(\hat{\theta})$ and the
maximum achievable or saturated likelihood
$\mathcal{L}(y)$, same as we do in Eq.~\eqref{eq:cminLR}.

Within the broader statistical community, Eq.~\eqref{eq:CminBaker}
is referred to as the Poisson `deviance' \citep[e.g., Eq.~(5.21) in][where the factor of 2 was omitted]{cameron2013}.
 A similar statistic to \eqref{eq:CminBaker} but without the terms $\hat{\mu}_i-y_i$ is referred to as the $G^2$
 statistic \citep{bishop1975}, whereas the omitted terms add to zero for models with an intercept.

 The terms `$C$ statistic', `C-stat' (and variants such as `cstat') and `Cash statistic' are often
 used interchangeably to refer to either \eqref{eq:CashCstat} with the data-dependent term, or \eqref{eq:CminBaker}
 without the data-dependent term. The data-dependent term is of course irrelevant for estimation, but needs to be
 taken into account for hypothesis testing. Further, in \texttt{XSPEC} and \texttt{SPEX} the use of the \texttt{cstat} statistic for spectral data with Poisson background means 
 the use of the \wmin\ statistic (or `wstat') of Eq.~\eqref{eq:wmin}, see Eq.~(B.7) in Appendix~B of the 
 manual \citep{xspecManual}. This use is discouraged for low-count data in the \texttt{SPEX} manual in favor of a parametric joint fit \citep{kaastra1996}.

 We refer to Cash, $C$, cstat  or \cmin\ as the statistic given by \eqref{eq:CminBaker}
 for a single source dataset, or by \eqref{eq:cmin} for the joint fit to source and background data, since it is
 unnecessary to consider the original definition with the data-dependent term. The \wmin, `wstat'  or \wstat\  is only meaningful for source and background datasets, and we use those terms interchangeably to refer to the statistic
 defined in \eqref{eq:wmin}.
}

\section{Goodness-of-fit statistics}
\label{sec:statistics}
The three methods of estimation are all based on the \ml\ criterion, although with
different hypotheses. It is therefore
necessary to investigate their asymptotic distributions for the purpose of hypothesis testing.
Equation~\eqref{eq:CminBaker}, which applies in slightly different forms in all three methods of estimation,
establishes the \cmin\ (joint fit and fixed-background) and \wmin\ (step-wise constant or non-parametric background) statistics as likelihood-ratio statistics. However, as we have extensively discussed in \cite{li2026},
these statistics do \emph{not} necessarily obey Wilks' theorem \citep[e.g.,][]{wilks1938}, 
except in the large-count limit. Therefore
they are not expected to follow exactly a chi-squared distribution with the appropriate number
of degrees of freedom in all data regimes. 

\subsection{General properties of $C$ statistics }
The main result of \cite{li2026} for the one-sample \cmin\ statistics is that they are approximately normally distributed in
the extensive data regime ($N \gg1)$.
Moreover, to the lowest degree of accuracy, the mean and the variance can be approximated via the sum of the respective moments for each
term of the statistic ($C_{\mathrm{min},i}$),  
\begin{equation}
\begin{cases}
    \E(\cmineq)\simeq \displaystyle\sum_{i=1}^N \E(C_{\mathrm{min},i})-p \quad \text{(mean)}\\[10pt]
    \Var(\cmineq) \simeq \displaystyle\sum_{i=1}^N \Var(C_{\mathrm{min},i}) \quad \text{(variance)},
\end{cases}
    \label{eq:cminMoments}
\end{equation}
where $p$ is the number of free parameters, or degrees of freedom, of the model.
The mean and variance for each data point can be calculated using simple approximations that were suggested by \cite{kaastra2017}
and \cite{bonamente2020} and hereafter referred to as the \emph{KB approximations},
where the $C_{\mathrm{min},i}$ are calculated for fixed values of the model parameters
as an approximation.~\footnote{Such approximation corresponds to considering the unconditional expectation of \cmin\ in the left-hand side, and that the $C_{\mathrm{min},i}$ statistics on the 
right-hand side are $C_{\mathrm{true}}$ statistics for fixed values of the parameters, in the terminology of \cite{li2026}.}
Newer and higher-accuracy results are provided in \cite{li2026}, although these approximations
are probably of sufficient accuracy for most astronomical applications.
 
 In particular, in the low-count regime (approximately less than one count per bin) the mean of $C_{\mathrm{min},i}$ can be substantially
lower than one according to the KB approximations, and somewhat larger than one
for an intermediate-count regime (few counts per bin), before approaching the chi-squared limit of one
in the large-count regime. Therefore, one can have an expected value of the statistic
that is substantially different from $N-p$, which is the expectation of a chi-squared distribution of same number of degrees of freedom.
Likewise, the variance of each term can be substantially different from 2, and therefore the total
variance can be less than the chi-squared value of $2 N$. 

In the large-mean limit, however, it remains true that the statistic is approximately $\cmineq \sim \chi^2_{N-p}$, same as for the usual chi-squared \gof\ statistic for Gaussian data.~\footnote{We use the symbol $\chi^2_{\nu}$ 
to denote a chi-squared random variable with $\nu$ degrees of freedom.} This can be seen 
by the fact that the Poisson distribution becomes equivalent to a normal distribution of same
mean and variance for large values of the Poisson mean \citep[e.g., Chapter~3 of][]{bonamente2022book}.

\subsection{\cmin\ statistics in the presence of background }
\label{sec:statisticsDistributions}

\subsubsection{The joint-fit \cmin\ statistic}
Given that the source and background region data are independent, the joint fit
corresponds to a total of $2N$ data points and $m+a$ adjustable parameters.
In the large-count limit, therefore, the joint fit statistic is distributed 
as the usual $\cmineq \sim \chi^2_{2N-(m+a)}$, under the null hypothesis that the 
source and background parametric models are accurate. The main feature of the joint
fit is that it uses a parametric model for the background region data, and therefore the resulting \gof\ 
statistic
tests the applicability of both models for the source and the background.

\subsubsection{The fixed-background \cmin\ statistic}
Similar to the previous case, the fit to the source data alone with a fixed background results in a \gof\
statistic that, in the large-count limit, is distributed as $\chi^2_{N-m}$. It is to
be noted that the null hypothesis under which the asymptotic distribution applies is that the fixed background is correct, which is not testable by this method of regression 
and it cannot be correct in general.  That assumption is different from the data model assumed in \eqref{eq:dataModel},
and therefore biases are expected when applying this method of regression to those data.

\subsubsection{The non-parametric-background \wmin\ statistic}
\label{sec:statisticsDistributionsWstat}
The case of non-parametric background is more complex, and it deviates from the
standard joint-fit method, as already described in Sec.~\ref{sec:wstatData}. First, the fit method
is such that the same data, i.e., both $S_i$ and $B_i$ datasets, are used twice in the regression:  to obtain the rMLEs
$\hat{b}_i(\theta)$ according to \eqref{eq:biHat}, and then in what is effectively a joint fit to
determine the MLE $\hat{\theta}$. 
Second, the background data are described by a non-parametric model, or by a model that has as many free parameters as
there are bins, while the source is described by the usual few $m$ parameters of interest.  This
is a situation when, especially for data of limited resolution (e.g., when $\beta\gg \mu$), the 
model introduces more \dof\ and it may lead to \emph{overfitting} \citep[see, e.g., Sec.~13.6 of][]{wasserman2010}. 

 Within the usual parametric framework,  the use of
\eqref{eq:biML} to obtain the rMLE estimates $\hat{b}_i(\theta)$ according to \eqref{eq:biHat}, followed by the usual (numerical) maximization
of the joint likelihood in \eqref{eq:profileLikelihood} suggest that the total
number of degrees of freedom is $2N-(m+N)=N-m$, same as for the case of the fixed background.
A similar conclusion would also result from considering the fit to the source data alone with a
model with the usual $m\ll N$ parameters for the source, considering $\hat{b}_i(\theta)$ according
to \eqref{eq:biHat} part of the model to the source data alone, if overfitting was
not a concern. 

As is discussed in the following section, however, this
simple method of estimating the number of degrees of freedom for \wmin\ is not accurate.
Moreover, the null hypothesis for \wmin\ includes asserting the accuracy of the non-parametric background model, which may not be correct in general.  The combination
of overfitting and inaccuracy of the null hypothesis will lead to substantial departures from the 
$\chi^2_{N-m}$ distribution, even in the large-count limit. This will be shown and discussed in more detail with the results of the simulations presented in Sec.~\ref{sec:MonteCarlo}.

\subsection{The number of degrees of freedom}
\label{sec:df}
It is usually tacitly assumed that the number of degrees of freedom of a parametric model
is equal to the number of adjustable parameters. Therefore, for $N$ independent measurements
and a model with $p$ parameters,
the number of \dof\ of the regression is $N-p$, as was implied in the foregoing.  This is 
however strictly true only for certain types of regression such as
the multiple linear regression \citep[e.g., see corollary to theorem 2 in][]{efron1986}, but not in
general \citep[e.g., see discussion in][]{janson2015, tibshirani2014}. Given the extensive use of
non-linear models in astronomy, the one-to-one correspondence of adjustable parameters
and degrees of freedom is often inaccurate \citep[see, e.g.][]{andrae2010a}.

A powerful and general-purpose method to estimate the number of \emph{effective}
degrees of freedom of a method of estimation or regression 
is provided by a theorem of \cite{efron1986}, which uses a generalization of 
Mallow's $C_P$ statistic \citep{mallows1973}.
This method
can in fact be immediately applied to our present case of Poisson regression \citep[as explained in Sec. 6 of][]{efron1986}. For data $y_i$ with common variance $\sigma^2$ and a method of estimation 
that leads to $\hat{\mu}_i=\mu(x_i;\hat{\theta})$, 
the (effective) number of degrees of freedom is defined by
\begin{equation}
    \df(\mu) = \dfrac{1}{\sigma^2} \sum_{i=1}^N \Cov(\hat{\mu}_i,y_i)
    \label{eq:df}
\end{equation}
as a corollary of Efron's theorem.
In general, $\df(\mu)$ can be smaller or larger than the number of
parameters used in the regression \citep[e.g.,][]{janson2015}, and it can in fact be used
for methods of estimation other than maximum-likelihood ones. Therefore, even
if the \wstat\ method is not a traditional joint-fit method for the reasons explained
in Sec.~\ref{sec:statisticsDistributionsWstat}, the considerations of this sections
apply.

The $\df(\mu)$ function is generally used in conjunction with the residual sum of squares 
\citep[and therefore the \chimin\ statistic;   e.g.,][for a convenient review]{janson2015}. More directly related to the Poisson data at hand,
it relates to the deviance or \cmin\ statistic itself, and to possible out-of-sample statistics
\cstar\ obtained by independent and identically distributed data $y^{\star}$ and the current
best-fit parameters. In particular,
\cite{efron1986} defines the `optimism' of a regression with $N$ data points and constant variance as 
a function $\omega(\mu)$ that, in the case of a Poisson regression, is given by
\begin{equation}
   N \cdot \omega(\mu)=\E(\cstareq) - \E(\cmineq)  = 2 \cdot \df(\mu)
   \label{eq:optimism}
\end{equation}
where \cstar\ is the \gof\ statistic using a new dataset $y^{\star}$ \citep[see Sec.~5 of][]{efron1986}.
The meaning is that each effective degree of freedom adds, on average, two units to the \gof\ statistic,
when a new dataset is used with the best-fit model, thus larger values of $\omega(\mu)$ indicate
a more optimistic estimate of the statistic with the measured data.~\footnote{Note that \cstar\ is different from
\ctrue, which is the \cstat\ using the true-yet-unknown parameter values, but for the
same data, see e.g. \cite{bonamente2025a}. Thus $\cstareq-\cmineq$ is \emph{not} the $\Delta C$ statistic that is customarily used to estimate confidence intervals on parameters of interest.} 

In the case of the multiple linear regression with $p$ independent regressors
it is in fact true that $\omega(\mu)=2 \cdot p/N$, e.g., $\df(\mu)=p=2$ for the usual linear regession with two parameters.
Moreover, for the same method of  regression the Wilks theorem guarantees that
asymptotically $\cmineq \sim \chi^2_{N-p}$.

For a general method of regression,
we therefore speculate that the results for the multiple linear regression can be generalized as follows:
\begin{conjecture}
The $\df(\mu)$ function of Eq.~\ref{eq:df} can be used to estimate  the
number of effective degrees of freedom for Wilks' theorem. Therefore, when applicable,
Wilks's theorem is modified to 
\begin{equation}
    \cmineq \sim \chi^2_{N-\df(\mu)},
    \label{eq:CminDf}
\end{equation} 
for a general method of regression.
\end{conjecture}


\begin{conjecture}
The substitution of $p$ with $\df(\mu)$ applies to Eq.~\eqref{eq:cminMoments},
    \begin{equation}
    \E(\cmineq)\simeq \displaystyle\sum_{i=1}^N \E(C_{\mathrm{min},i})-\df({\mu}).
    \label{eq:cminMomentsDf}
\end{equation}
\end{conjecture}

The two conjectures propose
that hypothesis testing with the \cmin\ and \wmin\ statistics is possible in all data regimes.  Empirical support for the
applicability of these conjectures is provided in remark~\ref{remark:conjecture} in Sec.~\ref{sec:efronSim}, based on simple
simulations with a constant model.
The  assumption of validity of Wilks' theorem is an important practical caveat, 
in particular requiring that $N$ is sufficiently large, and that the null-hypothesis model is properly specified. The latter condition, as 
discussed in Sec.~\ref{sec:statisticsDistributions}, includes the applicability of the background model.

It is therefore important to ascertain that the correct number of degrees of freedom of a model
is properly estimated. This issue is investigated in Sec.~\ref{sec:MonteCarlo} by means of numerical simulations.
It is important to point out that the Efron $\df(\mu)$ function is usually used to
quantify the degree of overfitting, rather than for an exact \gof\ estimate as is proposed in
the two conjectures. In the absence of an exact mathematical proof of their applicability, these generalizations
are therefore only plausible conjectures that are to be used as  empirical tools.

Another issue related to model estimation is that of structural parameter \emph{identifiability}, defined as different parameter values
leading to different models \citep[see, e.g., a recent review by][]{vannoort2024}. Moreover,
data of insufficient quality may lead to practical unidentifiability, which can be
diagnosed by a singular Fisher information matrix, among other criteria \citep[e.g.,][]{rothenberg1971}.
The simple constant models used in this paper clearly do not suffer from identifiability issues because of their extreme simplicity, but
this may become an additional concern for more complex source models.

\section{Results from numerical simulations}
\label{sec:MonteCarlo}
We performed of a series of Monte Carlo simulations aimed at
establishing the distributions of the \gof\ statistics of the three methods
and quantifying biases introduced by the fitting methods.

For this purpose we chose a simple constant model for both the source and the
background, so that $\mu_i(\theta)=\theta$, where $\theta>0$ is constant 
for all $x_i$, 
and $b_i=\phi$, also with $\phi>0$; we also assumed $t_B=t_S=1$ for convenience.
While this model may seem overly simplistic for 
many astrophysical applications, more complex models can
approximated by a constant over a narrow range of the independent variable. 
Moreover,
a constant model is such that Poisson data points have the same variance, and therefore
\eqref{eq:df} applies, leading to a direct use of Efron's theorem on the number of effective degrees of freedom. This simple model therefore aims at establishing the first-order features and biases of the
fit statistics, without being encumbered by the complications of more complex models. 

Results of the simulations for datasets with $N=100$ data points are shown in Table~\ref{tab:MonteCarlo}, with results
for $N=10$ and $N=1000$ being discussed in the appendix. We chose four values for
the parent mean, $\theta=0.1, 1, 10$ and $100$, which include representative values from the low-mean to 
the large-mean regime, and the same four values for the parent background model $\phi$. This choice
also lets us investigate extreme cases of source-dominated  ($\theta \gg \beta$) and background-dominated  ($\theta \ll \beta$) situations,
both common in astrophysical applications, along with intermediate cases, over a range of Poisson means.

\begin{table*}
    \centering
    \begin{tabular}{ll|lll|lll|lll}
    \hline
    \hline
\multicolumn{2}{c}{Intensity} & \multicolumn{3}{c}{$W_{\mathrm{min}}$ fit} & \multicolumn{3}{c}{$C_{\mathrm{min}}$ (joint) fit}&  \multicolumn{3}{c}{$C_{\mathrm{min}}$ (fixed back.) fit} \\ 
\hline
$\theta$ & $\beta$   &  stat. & bias & df($\mu$)  &  stat. & bias & df($\mu$)  &  stat. & bias & df($\mu$) \\ 
\hline
0.10 &  0.10 &76.3 $\pm^{9.6}_{8.6}$ & 0.83 $\pm^{0.44}_{0.41}$ & 35.54 & 114.8 $\pm^{13.1}_{12.7}$ & 0.00 $\pm^{0.50}_{0.50}$ & 1.96 & 81.2 $\pm^{11.0}_{8.8}$ & 0.84 $\pm^{0.47}_{0.43}$ & 0.98 \\ 
 1.00 &  0.10 &117.5 $\pm^{11.6}_{12.0}$ & 0.05 $\pm^{0.10}_{0.05}$ & 15.59 & 161.5 $\pm^{14.8}_{14.9}$ & 0.00 $\pm^{0.11}_{0.11}$ & 2.03 & 120.5 $\pm^{11.9}_{12.4}$ & 0.05 $\pm^{0.10}_{0.05}$ & 0.75 \\ 
 10.00 &  0.10 &100.6 $\pm^{15.0}_{14.4}$ & -0.00 $\pm^{0.03}_{0.03}$ & 2.05 & 146.3 $\pm^{17.9}_{17.0}$ & -0.00 $\pm^{0.03}_{0.03}$ & 2.05 & 101.5 $\pm^{15.1}_{14.5}$ & -0.00 $\pm^{0.03}_{0.03}$ & 1.01 \\ 
 100.00 &  0.10 &98.8 $\pm^{14.7}_{13.2}$ & -0.00 $\pm^{0.01}_{0.01}$ & 1.12 & 146.1 $\pm^{15.3}_{16.1}$ & -0.00 $\pm^{0.01}_{0.01}$ & 2.02 & 98.9 $\pm^{14.8}_{13.2}$ & -0.00 $\pm^{0.01}_{0.01}$ & 1.02 \\ 
 0.10 &  1.00 &145.8 $\pm^{17.7}_{16.4}$ & 5.36 $\pm^{1.34}_{1.08}$ & 34.13 & 229.0 $\pm^{16.7}_{16.0}$ & -0.00 $\pm^{1.40}_{1.40}$ & 1.94 & 177.9 $\pm^{20.4}_{19.1}$ & 5.36 $\pm^{1.29}_{1.09}$ & 0.09 \\ 
 1.00 &  1.00 &127.0 $\pm^{17.4}_{16.4}$ & 0.34 $\pm^{0.16}_{0.17}$ & 28.22 & 225.7 $\pm^{19.1}_{18.5}$ & 0.01 $\pm^{0.17}_{0.18}$ & 2.13 & 151.1 $\pm^{18.9}_{18.8}$ & 0.37 $\pm^{0.15}_{0.16}$ & 1.05 \\ 
 10.00 &  1.00 &101.6 $\pm^{15.4}_{13.3}$ & 0.00 $\pm^{0.04}_{0.04}$ & 8.81 & 213.9 $\pm^{20.2}_{18.3}$ & 0.00 $\pm^{0.03}_{0.04}$ & 2.09 & 109.3 $\pm^{16.2}_{14.7}$ & 0.01 $\pm^{0.03}_{0.04}$ & 0.95 \\ 
 100.00 &  1.00 &98.7 $\pm^{13.9}_{13.2}$ & 0.00 $\pm^{0.01}_{0.01}$ & 1.98 & 212.4 $\pm^{18.1}_{18.1}$ & 0.00 $\pm^{0.01}_{0.01}$ & 1.98 & 99.5 $\pm^{14.2}_{13.1}$ & 0.00 $\pm^{0.01}_{0.01}$ & 1.01 \\ 
 0.10 &  10.00 &102.0 $\pm^{15.2}_{13.8}$ & 0.10 $\pm^{4.53}_{1.10}$ & 49.96 & 202.5 $\pm^{20.9}_{20.3}$ & 0.10 $\pm^{4.30}_{4.40}$ & 1.92 & 196.0 $\pm^{29.2}_{25.7}$ & 9.83 $\pm^{4.33}_{3.89}$ & 0.96 \\ 
 1.00 &  10.00 &100.6 $\pm^{14.9}_{13.9}$ & 0.02 $\pm^{0.47}_{0.49}$ & 48.02 & 199.7 $\pm^{22.7}_{17.7}$ & 0.01 $\pm^{0.47}_{0.47}$ & 2.03 & 186.8 $\pm^{26.8}_{25.6}$ & 0.92 $\pm^{0.43}_{0.47}$ & 1.17 \\ 
 10.00 &  10.00 &99.6 $\pm^{15.5}_{13.1}$ & 0.00 $\pm^{0.05}_{0.06}$ & 34.24 & 199.7 $\pm^{21.3}_{20.0}$ & 0.00 $\pm^{0.05}_{0.06}$ & 1.96 & 146.9 $\pm^{22.5}_{20.2}$ & 0.05 $\pm^{0.05}_{0.06}$ & 1.14 \\ 
 100.00 &  10.00 &98.3 $\pm^{14.2}_{13.0}$ & -0.00 $\pm^{0.01}_{0.01}$ & 9.18 & 197.8 $\pm^{19.5}_{17.9}$ & -0.00 $\pm^{0.01}_{0.01}$ & 2.12 & 107.1 $\pm^{15.1}_{14.2}$ & 0.00 $\pm^{0.01}_{0.01}$ & 1.02 \\ 
 0.10 &  100.00 &99.0 $\pm^{15.4}_{12.8}$ & 0.13 $\pm^{14.21}_{1.13}$ & 50.32 & 198.3 $\pm^{20.8}_{18.2}$ & 0.10 $\pm^{14.32}_{13.10}$ & 1.98 & 196.8 $\pm^{29.7}_{25.7}$ & 10.40 $\pm^{13.60}_{11.40}$ & 0.88 \\ 
 1.00 &  100.00 &98.1 $\pm^{14.0}_{12.7}$ & -0.03 $\pm^{1.41}_{0.97}$ & 50.09 & 197.4 $\pm^{19.6}_{20.2}$ & -0.06 $\pm^{1.46}_{1.41}$ & 1.96 & 194.7 $\pm^{28.2}_{25.5}$ & 1.00 $\pm^{1.36}_{1.48}$ & 1.19 \\ 
10.00 &  100.00 &98.3 $\pm^{14.4}_{13.1}$ & 0.00 $\pm^{0.14}_{0.15}$ & 48.24 & 197.1 $\pm^{20.3}_{18.4}$ & 0.00 $\pm^{0.14}_{0.14}$ & 2.01 & 187.4 $\pm^{26.8}_{25.5}$ & 0.10 $\pm^{0.14}_{0.17}$ & 1.29 \\ 
 100.00 &  100.00 &98.6 $\pm^{15.1}_{14.2}$ & 0.00 $\pm^{0.02}_{0.02}$ & 34.13 & 198.8 $\pm^{19.2}_{19.6}$ & 0.00 $\pm^{0.02}_{0.02}$ & 2.08 & 147.6 $\pm^{22.6}_{21.2}$ & 0.01 $\pm^{0.02}_{0.02}$ & 1.08 \\ 
\hline
\hline
    \end{tabular}
    \caption{Results of the numerical simulations for constant source $\theta$ and 
    background $\beta$, for $1,000$ realizations and $N=100$ data points. The column `bias' is the fractional bias, and  `df$(\mu)$'  is the number of effective degrees of freedom of the model according
    to Eq.~\ref{eq:df}.}
    \label{tab:MonteCarlo}
\end{table*}

\begin{figure}
    \centering
    \includegraphics[width=0.95\linewidth]{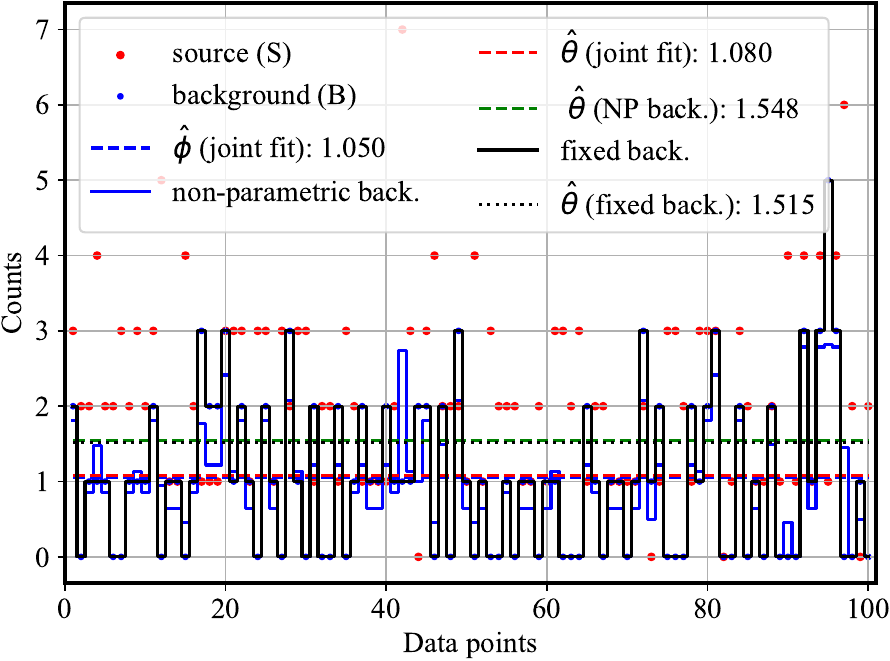}
    \caption{Sample data set and best-fit models for $\theta=\beta=1$ and $t_S=t_B=1$, with $N=100$. Data points for the source region (red) are drawn from a Poisson distribution with mean $\theta+\beta$, and
    for the background region (blue) from a Poisson distribution with mean $\beta$.}
    \label{fig:data11}
\end{figure}

A representative dataset for $\theta=\beta=1$ is shown in Figure~\ref{fig:data11}, in order to illustrate 
the salient features of the methods of regression of Poisson data with Poisson background. For the joint fit, the source and background region data
are fit independently to a constant model. For the fit with fixed background, the source data (red data points)
are fit to a model that consists of $\theta+B_i$, as per \eqref{eq:SiFixed}, so that the model
is always non-negative but it is forced to follow all the fluctuations of the background. For the \wstat\ method, first the step-wise constant background model is obtained
according to Eq.~\eqref{eq:biHat}, and then the model $\theta+\hat{b}_i(\theta)$ is fit to obtain $\hat{\theta}$
numerically; the blue step-wise curve represents the background $\hat{b}_i(\hat{\theta})$ for the best-fit
source level. A key feature of the \wstat\ method is that $\hat{b}_i(\theta)=0$ when $B_i=0$ and
for sufficiently low number of source counts $S_i$, as discussed after Eq.~\eqref{eq:biZero}.
This non-linear feature of the model results in the non-parametric background model often being pegged
at the level of 0 counts in bins with measured $B_i=0$.

\subsection{MLE constraints on best-fit values}
\label{sec:MLEconstraints}
Prior to the analysis of the results, we discuss constraints enforced by the \ml\ estimation method
for these constant-source models that are useful to interpret the numerical results.

The joint fit has MLE estimates given by 
\begin{equation}
    \begin{cases}
        \hat{\phi}=n_B/N \geq0,\\
        \hat{\theta}=n_S/N-\hat{\phi}
    \end{cases}
\end{equation}
where $n_B$ is the total number of counts
in the background region, and $n_S$ is the total number of counts
in the source region. Clearly, if $n_B \geq n_S$, which will occur due to Poisson fluctuations in
background-dominated data, then $\hat{\theta}\leq0$ is the MLE estimate, which is 
however not an acceptable value for a Poisson mean.

For the constant-background fit, the condition $\partial \cmineq/\partial \theta=0$ with \cmin\ given in Eq.~\eqref{eq:cminCB}, or equivalently using the Poisson likelihood based on the data model \eqref{eq:SiFixed},
leads to the condition
\begin{equation}
    N = \sum_{i=1}^N \dfrac{S_i}{\hat{\theta}+B_i},
    \label{eq:MLCBCondition}
\end{equation}
(for $t_B=t_S=1$) where $B_i$ is the fixed background count in the $i$--th bin. This equation
can be easily solved numerically for the MLE estimate $\hat{\theta}$. 

Moreover, it is also
useful to point out the
expression for \cmin\ given in \eqref{eq:cminCB} requires $\hat{\theta}\geq 0$ if there are bins where
$B_i=0$, a likely occurrence in low-mean data.

A simple analytic equation for the MLE estimate of $\theta$ in the case of non-parametric background cannot
be immediately given, due to the complexity of the fitting method. However, similar to the case above for the constant background, the \wmin\ statistic
in \eqref{eq:wmin} requires  $\hat{\theta}\geq 0$ for data where any of the $\hat{b}_i(\hat\theta)=0$,
which may occur when the measured background is $B_i=0$, as shown in Eq.~\eqref{eq:biHat}. Again, this is a likely occurrence in low-mean data, as also illustrated in Fig.~\ref{fig:data11}

\subsection{Goodness of fit}

\begin{figure*}
    \centering
    \includegraphics[width=0.48\linewidth]{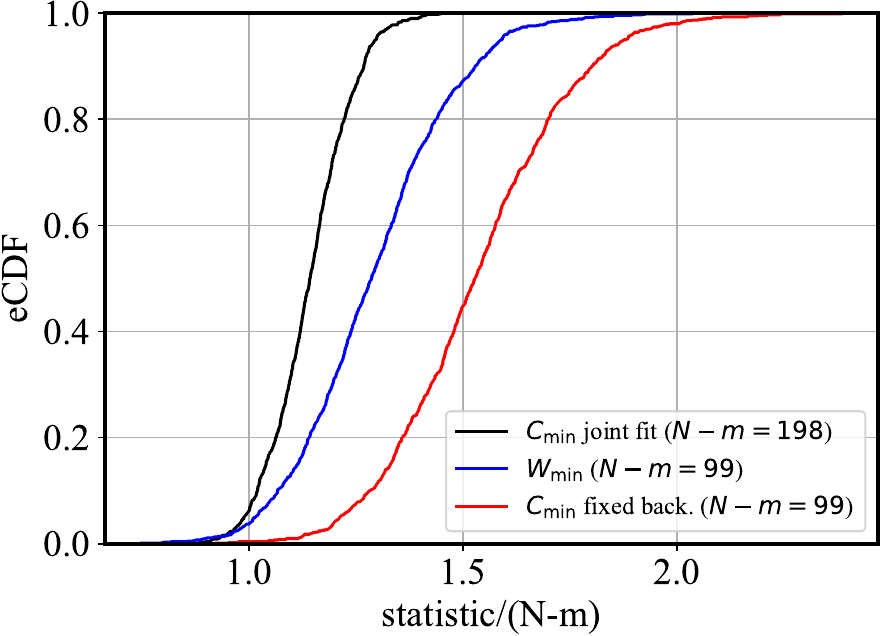}
    \includegraphics[width=0.48\linewidth]{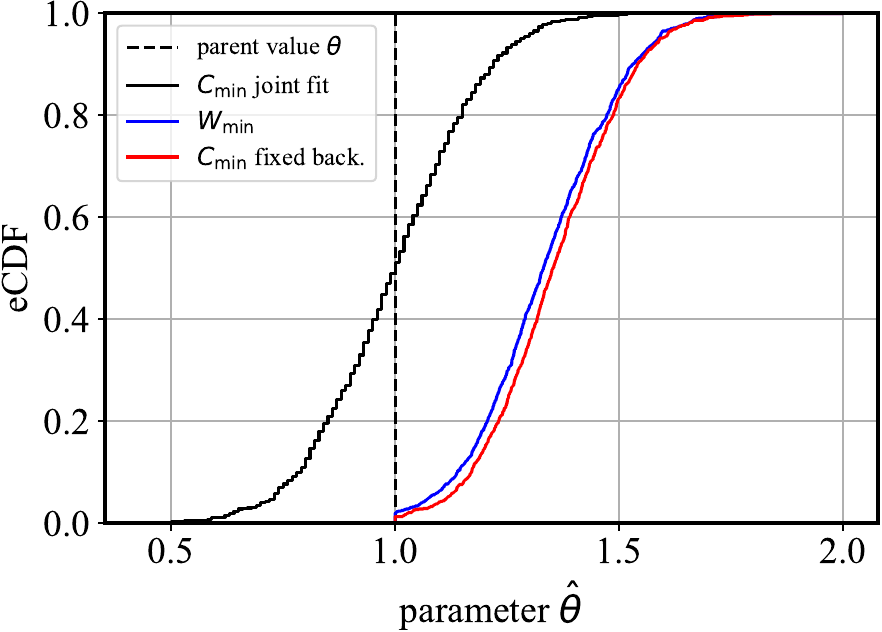}
    \caption{Experimental Cumulative Distribution Function (eCDF)  for the statistics and  best-fit parameters, based on 1,000 simulations with intensity $\theta=\beta=1$ and $N=100$ data points.}
    \label{fig:CDFbias}
\end{figure*}

The expectation and variance of the \cmin\ statistic for the fit of $N=100$ data points to a constant
model are reported in Table~\ref{tab:cmin}, for selected choices of the parent Poisson mean. For all parent means, the expectations based on
the KB approximations are very similar to those using the more accurate conditional moments of \cite{li2026},
and they differ significantly from the asymptotic values based on a chi-squared distribution, following
the key features already illustrated in \cite{bonamente2020}. For the variance, there is a good agreement 
between the two methods except at the lowest means, where the correction to the KB approximation
becomes dominant, as discussed in \cite{li2026}, to which we defer for further discussion of the
distribution of \cmin\ statistics. 
The results of Table~\ref{tab:cmin}, together with the effective number of degrees of freedom
from Table~\ref{tab:MonteCarlo}, can be used to
assess the quality of fit using the three methods of regression under consideration. The empirical cumulative distribution functions
(eCDF) for the \gof\ statistics, based on 1,000 simulations for a representative case with $\theta=\beta=1$, are shown in the left-hand panel of Fig.~\ref{fig:CDFbias}.

\begin{remark}[Accuracy of \cmin\ for joint fit]
The \cmin\ for the joint fit (with $2 N=200$ points and $2$ free parameters) follows the
theoretical expectations. For example, the case of $\theta=0.1, \beta=0.1$ corresponds to two
instances of the first line in Table~\ref{tab:cmin}, with an overall \cmin\ that is much smaller than the
asymptotic chi-squared case; or the case of  $\theta=10, \beta=0.1$ corresponds to the sum of
the results in the first and third line of the same table; etc. 
For background-dominated data, the source region has a parent mean of $\theta+\beta$, and so all cases 
correspond to large-mean Poisson data. The joint-fit results are therefore a sanity check that
the parent model is accurately reconstructed by the fit, and that there is agreement with
the expectation based on the theoretical results of \cite{li2026}.
Moreover, it is important to point out that the $\df(\mu)$ function consistently returns the expected value of 2, indicating that
the joint fit method indeed has two degrees of freedom according to the \cite{efron1986} theorem, as expected given the simplicity of the method of regression and the linearity of the model.     
\end{remark}

\begin{remark}[$\wmineq \leq \cmineq(\mathrm{FB})$]
\label{remark:Wmin<Cmin}
The \wmin\ statistic is always lower than the \cmin\ statistic with fixed background.
This was expected, 
given the choice to optimize a non-parametric background model for the \wstat\ method according to Eq.~\eqref{eq:biML}. In the case of source-dominated data ($\theta \gg \beta$) the two statistics are similar,
see e.g. the case of $\mu=100$ and $\beta=0.1$. In the case of background-dominated data ($\theta \ll \beta$), on the other hand,
$\wmineq \ll \cmineq$. This is in part due to the fact that the \wstat\ method has a
significantly larger number of effective degrees of freedom, compared to the \cstat\ with fixed background.
For example, for $\theta=1$, $\beta=10$  the \wstat\ method has significantly more
degrees of freedom than the single degree of freedom for the constant source model used in the \cstat\
method with fixed background. This is further addressed in remark~\ref{remark:overfitting}.    
\end{remark}

\begin{remark}[Model mis-specification for \cmin(FB)]
In the source-dominated limit, the fixed-background \cstat\ method has a \cmin\ statistic that is consistent with the 
expectation for a model with one degree of freedom (and for the appropriate mean, see Table~\ref{tab:cmin}). For background-dominated data, however, the fixed-background method
has a larger \cmin\ value than expected, see e.g. the case of $\theta=1$, $\beta=10$. This is simply due to  model mis-specification, i.e., the fact that the data are not consistent with the  null-hypothesis model (i.e., that the true background is the fixed background), leading to a poor \gof\ statistic.    
\end{remark}

\subsection{Biases in the reconstruction of source parameters}

A summary of the biases introduced by the fitting methods is presented in 
Table~\ref{tab:MonteCarlo}. 
The right-hand panel of Figure~\ref{fig:CDFbias} shows the distribution of biases for the three
methods, for the case of $\theta=\beta=1$. The figure illustrates clearly how the joint fit is unbiased, and
the biases introduced by the non-parametric and fixed-background fits.


\begin{remark}[Unbiasedness of joint-fit]
    The joint-fit method is successful at reconstructing the source parameter, even in background-dominated
    cases, see red curve in right-hand panel of Figure~\ref{fig:CDFbias}. A small bias may be present
    in the most extreme cases, e.g., $\theta=0.1, \beta=100$, e.g., when the background is 1,000 times the source level. This is largely due to the fact that only a few source counts are
    present in the simulations, given the choice of $N=100$ data points.  We have shown that this bias 
    disappears for larger values of $N$ that ensures a large number of source counts, using additional simulations described in the appendix.
\end{remark}

Both non-parametric and fixed-background fits are significantly biased, especially in the low-mean regime $\theta \simeq \beta \leq 1$, with the behavior in the right-hand panel of Fig.~\ref{fig:CDFbias} for $\theta=\beta=1$ being found also for smaller values of the two parameters
(see appendix). 
In particular, we find that $\hat{\theta} \geq \theta$ for all simulated data in this regime, 
in contrast with the
joint fit case where the MLE estimate is allowed to vary around the parent value. For the constant-background case, this behavior can be explained quantitatively by the MLE requirement of Eq.~\eqref{eq:MLCBCondition}. In fact, in the low-count limit there will be bins with $B_i=0$, and a value of $\hat{\theta} \leq \theta$ would give rise to large contributions $S_i/\hat{\theta}$, with $S_i \sim \Poiss(\theta+\beta)$, that would lead to a violation of the MLE constraint. 

For the non-parametric case, it can be shown that a similar constraint to Equation~\eqref{eq:MLCBCondition} also applies, if we assume that the $\theta$-dependence of $\hat{b}_i$ is ignored (see Appendix~\ref{appB}). Therefore the same considerations also apply
to the non-parametric background to explain the strong bias in the low-mean regime.

\begin{remark}[Biases with \wmin\ and \cmin(FB) in low-mean data]\label{remark:WLowMean}
    For background-dominated  and  low-mean data ($\theta \leq \beta \leq 1$) the non-parametric
    and fixed-background methods
    introduce significant biases in the reconstruction of the source parameter $\theta$, 
    in the direction of a systematic \emph{overestimation} of the source intensity. 
    For a given background level, these biases are
     reduced as the source mean increases, as expected. These two methods are therefore asymptotically
     unbiased as $\theta/\beta \gg 1$.  In both cases, the biases are a result of
     model mis-specification, i.e., the use of a background model that is different from that of the data-generating process.
\end{remark}

\begin{table}[!b]
    \centering
    \begin{tabular}{l|ll|ll|ll}
    \hline
    \hline
    Mean & \multicolumn{2}{c}{KB Approx.} & \multicolumn{2}{c}{Conditional mom.} & \multicolumn{2}{c}{$\chi^2$ dist.}\\
      $\mu$   & \footnotesize $\E(\cmineq)$ & \footnotesize $\Var(\cmineq)$ &\footnotesize $\E(\cmineq)$ &\footnotesize $\Var(\cmineq)$\\
    \hline
    0.1     & 48.2 & 82.9 & 49.7 & 4.2   & 99 & 198  \\
    0.3 & 75.5 & 68.3 & 76.8 & 18.3 &  99 & 198\\
    1.0     & 112.7 & 137.8 & 113.6 & 136.9 & 99& 198 \\
    3.0     & 107.5 & 240.6 & 108.4 & 240.1 & 99    & 198   \\
    10      & 99.8 & 208.4 & 100.8 & 208.4 & 99 & 198 \\
    100     & 98.2 & 200.7 & 99.2 & 200.7  & 99& 198 \\
    \hline
    \hline
         
    \end{tabular}
    \caption{Expectation and variance of \cmin\ based on the KB approximations \citep{kaastra2017,bonamente2020} and on the \cite{li2026} conditional means, for a dataset with $N=100$ Poisson data points with parent mean $\mu$ and fit to a constant model.}\label{tab:cmin}
\end{table}

\subsection{The {Efron} function and hypothesis testing}
\label{sec:efronSim}
For the two methods of regression with \cmin\ (joint fit and fixed background),
the Efron $\df(\mu)$ function returns, on average,  the values of approximately 2 and 1 respectively.
These values are consistent with the `classical' view that the number of parameters equals the number of degrees of freedom of the
regression, given that $m+a=2$ parameters were used in the joint fit,
and just $m=1$ parameter for the fixed-background method. 
This means that the \cmin\ methods, at least for the constant models investigated
in this paper, are sufficiently simple that each parameter accounts for one effective degree of
freedom, in the sense of Efron's theorem of Eq.~\eqref{eq:df}. 

\begin{remark}[Empirical validation of conjectures \eqref{eq:CminDf} and \eqref{eq:cminMomentsDf}]
\label{remark:conjecture}
    As a result, for both the joint-fit method
and the fixed-background method, hypothesis testing can be performed using Equations~\eqref{eq:cminMoments}
with $p$ set at the number of adjustable parameters, respectively $p=m+a=2$ for the joint fit and
$p=m=1$ for the fixed background, when the source model is constant. In the large-mean limit, 
the  $\cmineq \sim \chi^2_{N-p}$ approximation can be used.  
This corresponds
to using the two conjectures in Sec.~\ref{sec:df}, namely Equations~\eqref{eq:CminDf} and \eqref{eq:cminMomentsDf}, and therefore these findings for the simple constant model, and in the large-mean limit, lend support to the two conjectures.
Failure of the null-hypothesis
test 
would therefore call into question its assumptions, which for the constant-background case includes
the applicability of the fixed level of background.
\end{remark}

The situation is more complex for the \wstat\ method.
 In  Sec.~\ref{sec:statisticsDistributions} we discussed that,
following the `classical' view, 
one would expect the
\wmin\ statistic to have $N-m$ degrees of freedom, with $N=100$ and $m=1$, same as for the
constant-background case.
The simulations are in stark disagreement with that view.
\begin{remark}[Overfitting and $\df(\mu) \gg p$ with \wmin]
\label{remark:overfitting}
    The simulations show 
that $\df(\mu) \gg p=1$, except in cases
of source-dominated data where we recover the result $\df \simeq 1$, see e.g.
the case of $\theta=100, \beta=0.1$ in Table~\ref{tab:MonteCarlo}.
The large value of $\df(\mu)$, especially for background-dominated data, can be seen
as a form of \emph{overfitting}, whereas the background model is allowed to (erroneously) follow
the Poisson fluctuations, albeit not exactly as in the fixed-background case,
for the purpose of likelihood maximization, see Eq.~\eqref{eq:biML}. As a result, the fixed-background regression
does not suffer from the problem of overfitting, although it is biased by the choice of
adding the (incorrect) fixed background to the model for the source region data.
\end{remark}

\begin{remark}[Hypothesis testing with \wmin]
\label{rmk:wminTesting}
    For the \wstat\ method, hypothesis testing can  be tentatively performed  using the two conjectures of Sec.~\ref{sec:df}. However,
    care needs to be used in evaluating the $\df(\mu)$ function for the data at hand first,  which can be substantially larger than $m$ for data of limited resolution. Same as for
    the earlier remark, failure of the \gof\ test should be used to call into question the assumption
    of the non-parameteric model.
\end{remark}

Following remark~\ref{rmk:wminTesting}, we now focus on examining in more detail 
the three cases in Table~\ref{tab:MonteCarlo} with $\theta/\beta=0.1$, which are representative of background-dominated data, from the low-count to the large-count limits. 

The joint fit can consistently reconstruct the source intensity accurately, 
and with an acceptable \gof\ statistic ($\cmineq/(2N-2) \simeq 1$). 
Naturally, there is a statistical uncertainty in the parameter estimates for a given regression. 
The considerations in Sec.~\ref{sec:MLEconstraints} lead immediately to $\E(\hat{\theta})=\theta$, thus showing analytically 
that the method is in fact unbiased, and moreover $\Var(\hat{\theta})=(\theta+\beta)/N$. 
For example, for $N=100$ and 
$\theta=0.1, \beta=1.0$ one expects to measure a best-fit source parameter of approximately $0.1\pm 0.15$,
for an expected bias of $0.0\pm1.5$, which is the same as the simulation-based bias $0.00\pm1.5$ in Table~\ref{tab:MonteCarlo}.

On the other hand, the \wstat\ method has a significant bias in the low-count regime, and it becomes
unbiased only in the large-count regime (e.g., $\theta=10, \beta=100$). Moreover, the \wmin\ statistics
are substantially larger than the expectations according to both \eqref{eq:cminMoments} or the
chi-squared approximations, since $\df(\mu) \gg 1$. For example, consider the 
case of $\theta=1,\beta=10$ and the median value of $\wmineq=100.6$.
The regression has an estimated $\df(\mu)\simeq 48$ degrees of freedom, and not just one according to the classical view
of degrees of freedom. Therefore $\E(\wmineq) \simeq 100-48=52$ (according to the chi-squared approximation)
or $\simeq 100.8 - 48 \simeq 52.8$ according to the \cite{li2026} approximations, both much larger
than the average measured value. 
A poor fit can be naturally explained with the inaccuracy of the non-parametric model, given that a constant parent model was used.

Finally, the constant-background case has larger biases than the \wstat, with a 10\% bias
even in the large-mean case of background-dominated data ($\theta=10, \beta=100$), and with
a larger-than-expected \cmin\ statistic in all three cases.  Its failure is again explained 
by the model mis-specification introduced by the fixed background.

\section{Discussion and Conclusions}
\label{sec:discussion}
This paper has investigated three popular methods for the regression of Poisson data in
the presence of Poisson background, which is a common occurrence in many astrophysical data analysis situations.

The natural method of regression for these data is the joint fit to a parametric model for
both the source and the background data. The method is unbiased, and it provides
hypothesis testing via the usual \cmin\ statistic, which is asymptotically distributed
as a chi-squared variable in the large-mean limit, and otherwise normally distributed according to the KB approximations discussed in Eq.~\ref{eq:cminMoments} \citep{kaastra2017,bonamente2020} or the more accurate \cite{li2026} approximations. This is the recommended method of analysis.

The \emph{wstat} method that uses a non-parametric model for the background can only be used in the limit
of large-mean Poisson data, as discussed in remark~\ref{remark:WLowMean} in Sec.~\ref{sec:MonteCarlo}.
The method, in fact, is systematically  biased to \emph{overestimate} the source intensity,
as already found by \cite{vianello2018}, due to the use
of a model that is generally mis-specified. The method becomes progressively unbiased in the limit of source-dominated data, i.e., as the background becomes
negligible. Therefore it can be used only when the background is a small fraction of the source intensity. 

When complex multi-parametric models are considered for either 
the source or the background, care must be exercised in assessing the correct number of degrees of freedom, which may
 be different from the number of free parameters. This
task can be accomplished by means of the Efron $\df(\mu)$ function, Equation~\eqref{eq:df},
which can  be estimated by means of simple numerical simulations of the type conducted for this paper.
Notice how
the equation assumes homoscedastic data and thus a uniform variance, which may not be appropriate in many settings when Poisson data with  a large dynamical range are present. In those cases, a generalization of Equation~\eqref{eq:df} to heteroscedastic data can be used instead,
by means of weights for each covariance term \citep[see, e.g.,][]{luan2022measuringmodelcomplexityheteroscedastic}.
If an estimate of the correct number of degrees of freedom is unavailable for a specific
regression, the data analyst should
at least enforce an empirical redundancy in the hypothesis testing process, since in general one expects $\df(\mu) \geq m$, with $m$ the number of parameters of the source model. 

Using a fixed background in the regression is simply too coarse  a method of regression in most cases, and it exacerbates the
problems highlighted with the \wstat\ method. A fixed background can only be used in large-mean and
source-dominated data, when the background is effectively negligible.

It is possible that a more judicious use of non-parametric models for the background
may ameliorate the issues identified in this paper with the \wstat\ method. In particular, 
 a non-parametric model of the background using the type of `adaptive rebinning'
commonly used in Bayesian blocks analysis of light curves \citep[e.g.,][]{scargle2013}, may be a
suitable avenue for modeling Poisson-count spectra of the type considered in this paper. Within such 
framework, the source would continue to have a parametric model applied to the original source spectrum, but the background would be rebinned and thus modeled
non-parametrically but without the strong non-linearities of the \wstat\ method.

One final comment is reserved for the rebinning of the data. In high-energy astrophysics data bin sizes are often arbitrary, since most instruments collect individual photon events. The only key requirement for an accurate statistical analysis
is the independence among Poisson measurements, as enforced by the resolution of the instrument, although other considerations may also come into play \citep[e.g.,][]{kaastra2016}. The oft-used prescription to `rebin until $\geq 10$' counts and use of Gaussian
statistics for Poisson data \citep[e.g.,][]{bevington2003}, is generally not recommended.
First, rebinning reduces the resolution of the data, and it is therefore not practical for many data analysis tasks
such as, e.g., the search for narrow emission of absorption lines \citep[e.g.,][]{spence2023,bonamente2025c}. Second,
even in the large count-per-bin limit, the use of Gaussian statistics leads to biases in the parameter estimation,
due to the incorrect estimate of the data variance by the number of detected counts \citep[e.g.,][]{humphrey2009}.

\appendix
\section{Simulations with $N=10$ and $N=1000$ data points}
\label{appA}
Two sets of simulations 
similar to those in Table~\ref{tab:MonteCarlo}, but for $N=1,000$ and $N=10$ datapoints, were also performed. These simulations
show the same general features for the bias, and the distribution of the \gof\ statistics, and the
number of effective degrees of freedom, as for $N=100$ data points, but with different levels of statistical noise. These additional simulations show that there are substantially larger values of $\df(\mu)$ compared to  $m=1$ for the \wmin\ method, and values consistent with respectively $m=1$ and $m+a=2$ for the two \cmin\ statistics with fixed and adjustable background, respectively.  These additional simulations therefore confirm the findings
for $N=100$, and suggest they apply to both sparse and extensive data.

\section{Analysis of selected simulations}
\label{appB}
\begin{figure*}
    \centering
    \includegraphics[width=0.48\linewidth]{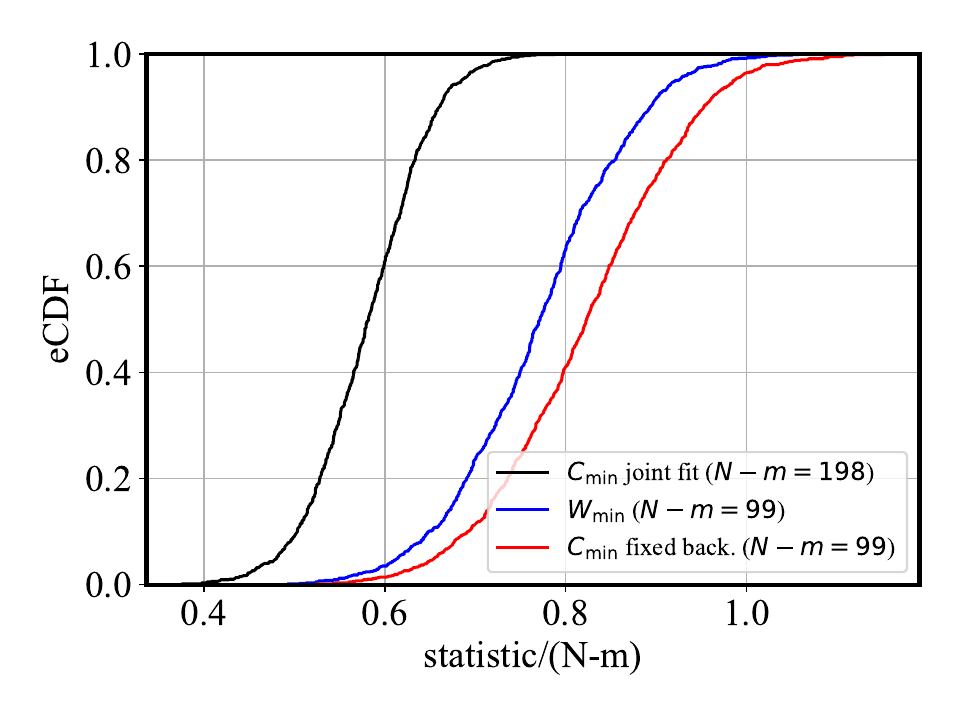}
    \includegraphics[width=0.48\linewidth]{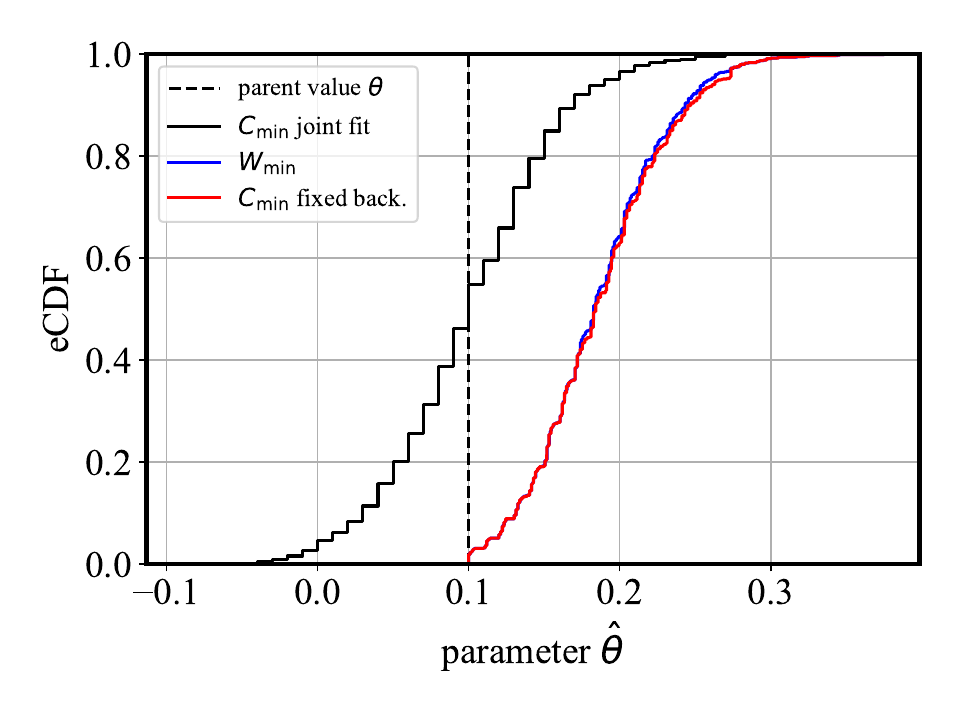}
    \includegraphics[width=0.48\linewidth]{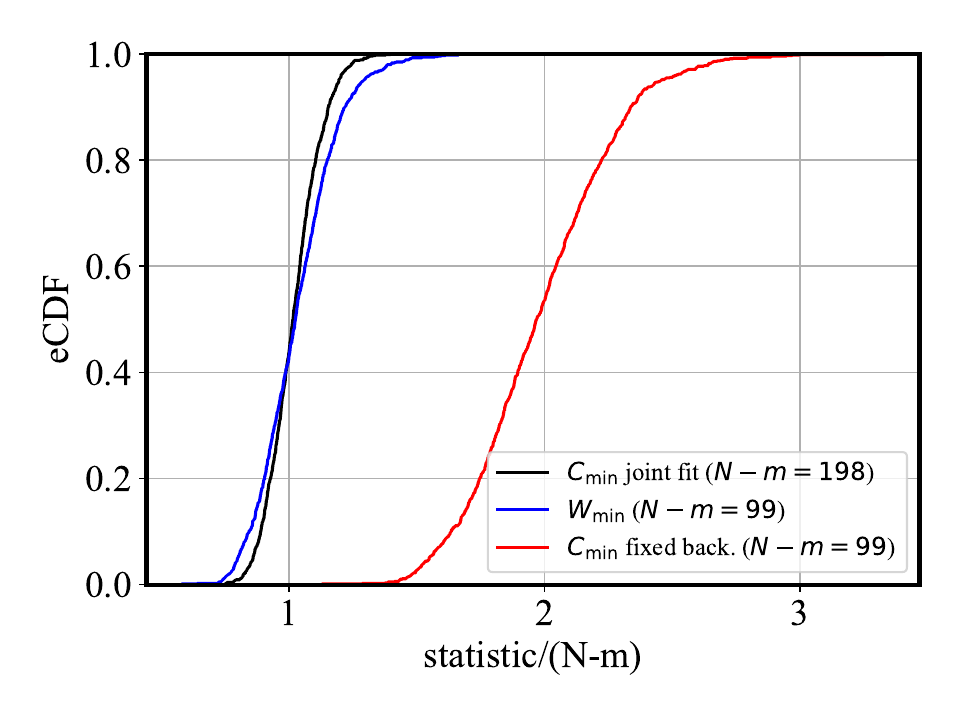}
    \includegraphics[width=0.48\linewidth]{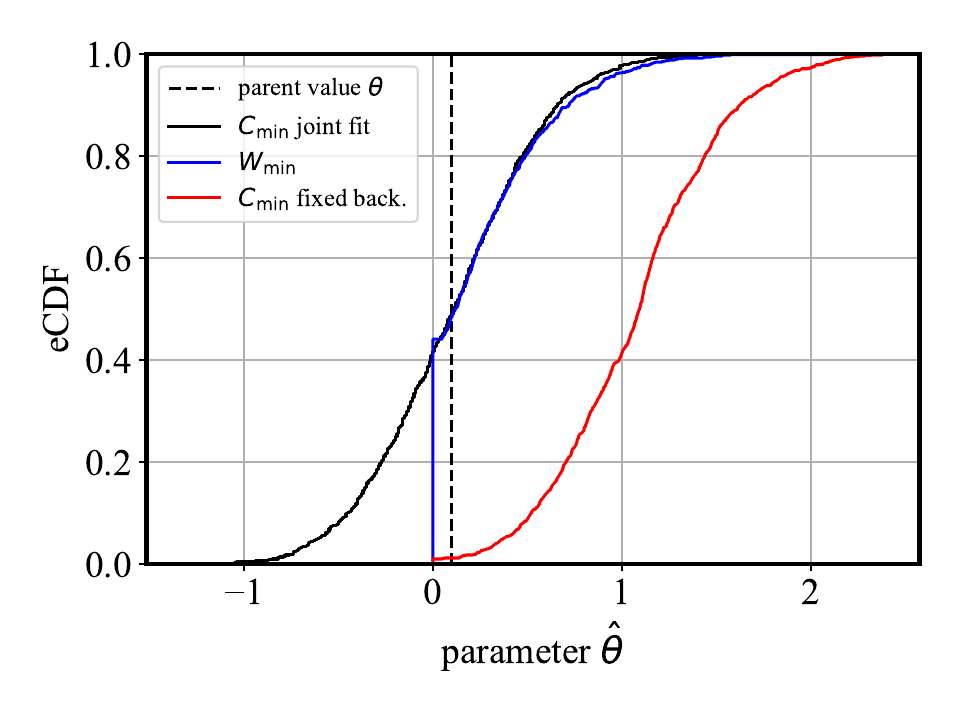}
    \caption{(Top): eCDFs for the simulations with $N=100$, $\theta=\beta=0.1$. (Bottom):  eCDFs for the simulations with $N=100$, $\theta=0.1, \beta=10$}
    \label{fig:theta0.1beta0.1N100}
\end{figure*}

Results of the simulations with with $\theta=\beta=0.1$ and $N=100$ are shown in the top row of Figure~\ref{fig:theta0.1beta0.1N100}, and those for $\theta=0.1, \beta=10$ and $N=100$ in the bottom
row. 

For the case of $\theta=\beta=0.1$, which is representative of low-mean data with comparable
source and background intensities, the joint fit is unbiased. The distribution of \cmin\ values 
for the joint fit is consistent with Equation~\ref{eq:cminMoments}  and expectations according to 
 Table~\ref{tab:cmin}. In this low-mean regime, $E(\cmineq)/N  \simeq 0.5$ for data with mean 0.1 (background data)
 and somewhat larger, albeit smaller than 1, for data with mean 0.2 (source data). The eCDF in the top row
 of Figure~\ref{fig:theta0.1beta0.1N100} shown a median value of $E(\cmineq)/N \simeq 0.55$, in accord with
 those theoretical expectations as presented in \cite{li2026}.
 Both fits with non-parametric and fixed backgrounds are significantly biased, as shown in the right panel,
 and with $\wmineq < \cmineq$, as shown in the left panel (this is also true for each of the 1,000 datasets individually). 

The case of $\theta=0.1$ and $\beta=10$ is representative of background-dominated data. The joint fit remains
unbiased, and the fixed-background is strongly biased; both have near-normal distributions of the
respective \gof\ statistics, see bottom-right panel of Figure~\ref{fig:theta0.1beta0.1N100}.
The non-parametric background, on the other hand, shows a more peculiar behavior, whereas the method does not allow negative values of $\hat{\theta}$, and the eCDF is accordingly truncated at $\hat{\theta}=0$. 
This behavior can be explained as follows.

In the background dominated-data regime, $S_i,B_i\gg1$
and $\hat{\theta}\ll1$. According to Eq.~\eqref{eq:biHat},
it is possible to show that
\begin{equation}\dfrac{d\hat{b}_i(\theta)}{d\theta}=
\dfrac{1}{2}\left(-1+\dfrac{\theta-(S_i-B_i)/2}{\sqrt{(\theta-(S_i-B_i)/2)^2 +S_iB_i}}\right) \simeq -\dfrac{1}{2}
\end{equation}
and accordingly the approximate solution of the score equation, 
if $\hat{\theta}/\hat{b}_i \ll 1$, is
\begin{equation}
    \hat{\theta} \simeq \dfrac{\sum S_i/\hat{b}_i - \sum B_i/\hat{b}_i}{\sum S_i/\hat{b}_i^2} \gtrsim  0
\end{equation}
since on average $S_i \geq B_i$. This explains the effective truncation of the best-fit values near $\hat{\theta}=0$, which did not occur for the joint fit.

\begin{figure*}
    \centering
    \includegraphics[width=0.48\linewidth]{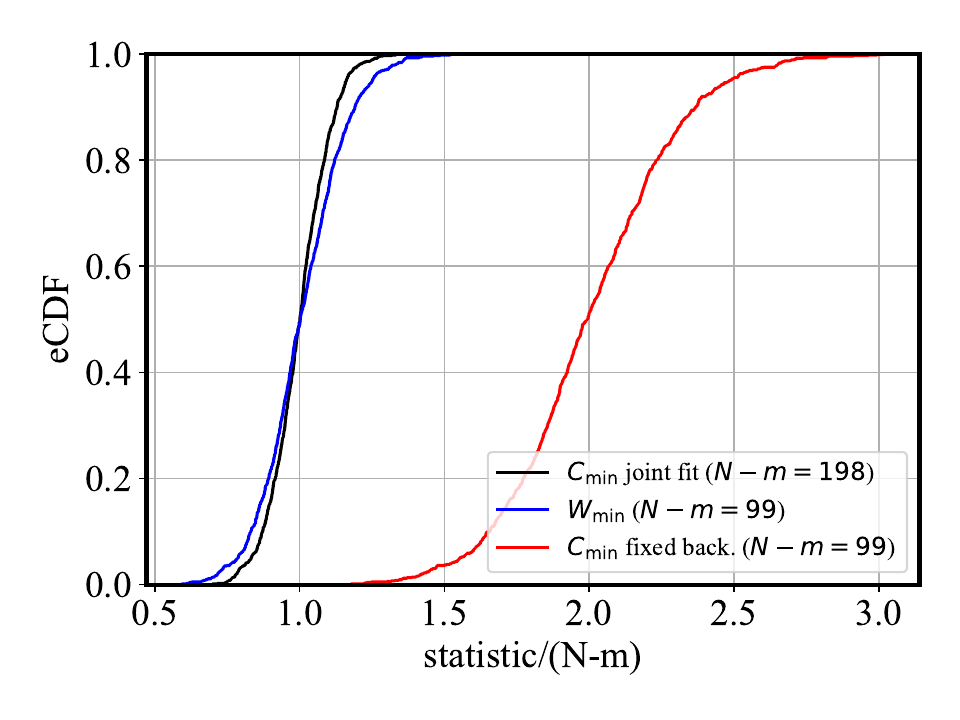}
    \includegraphics[width=0.48\linewidth]{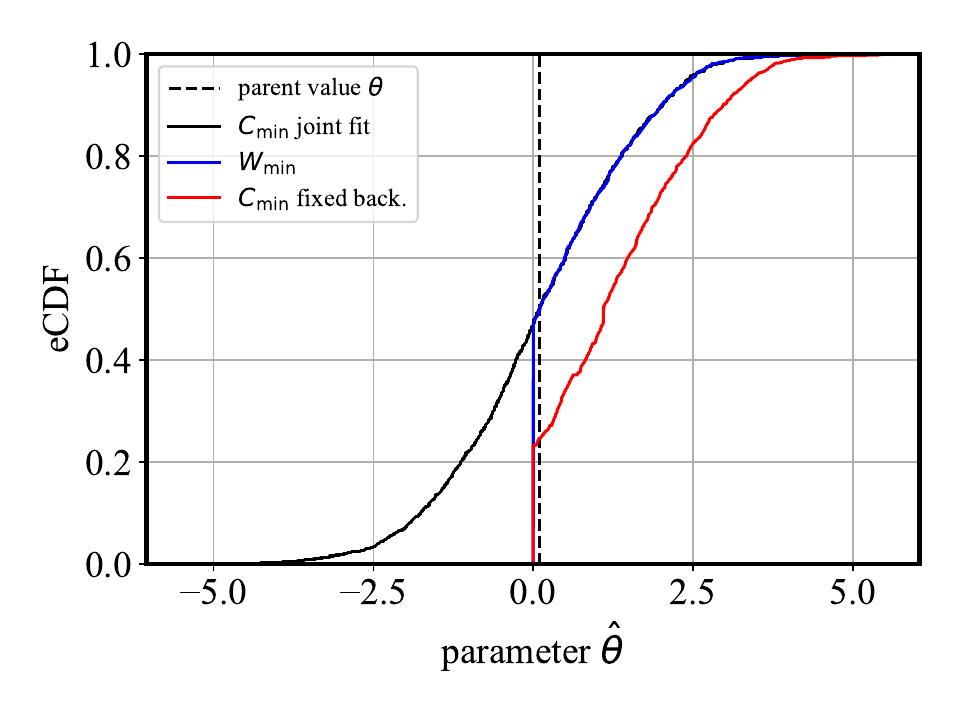}
    \caption{eCDFs  for the most extreme case of $\theta=0.1$, $\beta=100$, $N=100$.}
    \label{fig:theta0.1beta100}
\end{figure*}

A similar non-linear behavior for the distribution of $\hat{\theta}$ may occur for the fixed-background case, for background dominated-data, as illustrated in the more extreme case of  $\theta=0.1$, $\beta=100$, $N=100$, see Fig.~\ref{fig:theta0.1beta100}. In fact, in this regime, the MLE constraints of Eq.~\eqref{eq:MLCBCondition} lead to
\begin{equation}
    \hat{\theta} \simeq \dfrac{\sum S_i/B_i -N}{\sum S_i/B_i^2}.
\end{equation}
Since on average $S_i/B_i\geq1$, the method is expected to return $\hat{\theta}\geq 0.$

\bibliographystyle{aasjournal}
\bibliography{math,max,urls}
\end{document}